\documentclass[%
 reprint,
 amsmath,amssymb,
 aps,
]{revtex4-2}
\usepackage{lipsum}
\usepackage{graphicx}
\usepackage{dcolumn}
\usepackage{bm}

\begin{document}

\preprint{APS/123-QED}

\title{Entanglement-assist cyclic weak-value-amplification metrology}

\author{Zi-Rui Zhong}
\author{Xia-Lin Su}
\author{Xiang-Ming Hu}
\author{Qing-Lin Wu}%
\email{qlwu@ccnu.edu.cn}
\affiliation{%
\emph{Department of Physics, Central China Normal University, Wuhan 430079, China}
}%

\date{\today}

\begin{abstract}

Weak measurement has garnered widespread interest for its ability to amplify small physical effects at the cost of low detection probabilities. Previous entanglement and recycling techniques enhance postselection efficiency and signal-to-noise ratio (SNR) of weak measurement from distinct perspectives. Here, we incorporate a power recycling cavity into the entanglement-assisted weak measurement system. We obtain an improvement of both detection efficiency and Fisher information, and find that the improvement from entanglement and recycling occur in different dimensions. 
Furthermore, we analyze two types of errors, walk-off errors and readout errors. The conclusions suggest that entanglement exacerbates the walk-off effect caused by recycling, but this detriment can be balanced by proper parameter selection. In addition, power-recycling can complement entanglement in suppressing readout noise, thus enhancing the accuracy in the measurement results and recovering the lost Fisher information. This work delves deeper into the metrological advantages of weak measurement.

\end{abstract}

\maketitle


\section{Introduction}

A quantum measurement commonly pertains to von Neumann model, which works by projecting a system onto an eigenstate of the system observable. This process necessitates a strong coupling between the system and the probe to ensure the measurement accuracy, with the measurement outcomes typically aligning with the eigenvalues of the system. 
Weak measurement, initially proposed by Aharonov, Albert and Vaidman in 1988 \cite{PhysRevLett.60.1351}, is a distinctive form of quantum measurement. It characterized by a weak coupling between the system and the probe, enabling the pointer observable to significantly surpass the range of eigenvalues of the system via postselection \cite{steinberg2010}. 
Generally, the large shift of pointer can approximately exhibit linear correlation with the amplification of the weak interaction, a phenomenon called weak-value-amplification (WVA) effect. Consequently, it has been found useful in observing small physical effects such as the spin Hall effect \cite{science.1152697,PhysRevA.85.043809,PhysRevLett.109.013901,Bai:20}, optical beam deflection \cite{PhysRevLett.102.173601,PhysRevA.80.041803}, phase shift \cite{PhysRevA.82.063822,PhysRevLett.111.033604,Qiu2017}, velocity\cite{Viza:13}, Goos-Hänchen shift \cite{Santana:16}, temperature \cite{Egan:12,10.1063/1.5027117}, angular rotation \cite{PhysRevLett.112.200401,DELIMABERNARDO20142029,Fang:21} and magnetic resonance \cite{Qu2020}. Some interesting phenomena have also been reconsidered in the literature \cite{PhysRevLett.68.2981,AHARONOV2002130,PhysRevLett.102.020404,Yokota_2009,YAharonov_1991,Ravon_2007,RESCH2004125,Aharonov_2013,denkmayr2014,DUPREY20181,aharonov2021,pan2020}. 

An entire weak measurement usually contains three stages: preparation of the system and pointer, weak coupling between them, and postselection of the system. In order to amplify small physical quantities, it is essential to set the pre- and post-selected states nearly orthogonal, thereby enlarging the weak value. However, a large amplified weak value corresponds to low postselection efficiency, resulting in a waste of resources and Fisher information of the measurement if discarding the failed postselection part. In terms of measurement precision, the loss entirely counterbalance the gain from WVA effect. This balance consideration has caused debates on whether weak measurement outperforms the conventional measurement \cite{PhysRevA.85.062108,PhysRevA.85.060102,PhysRevX.4.011031,PhysRevA.88.042116,PhysRevLett.107.133603,PhysRevLett.114.210801,Jordan2015,PhysRevLett.112.040406,PhysRevA.106.022619,PhysRevA.102.042601,PhysRevA.91.032116,PhysRevX.4.011032,PhysRevA.91.062107,PhysRevA.96.052128}. It is clear that post-selection itself cannot enhance the precision, but some certain conditions can make weak measurement optimal. Actually, it has been proven that weak measurement can outperform the conventional measurement in the presence of detector saturation \cite{PhysRevLett.118.070802,PhysRevLett.125.080501}. In addition, weak measurement has been pointed out to suppress the technique noise in some cases and can outperform conventional measurement by several orders of magnitude via imaginary weak value \cite{PhysRevLett.105.010405,PhysRevA.85.060102,PhysRevA.102.042601,PhysRevX.4.011031,RevModPhys.86.307,PhysRevLett.132.043601}. Also, significant improvements in precision metrology, with sensitivity enhanced by several orders of magnitude, can be accomplished through a technique named dual WWA \cite{PhysRevA.100.012109}. 
Moreover, when measuring the distribution of postselection process, Heisenberg-scaling precision can be achieved by utilizing seeming classical resources \cite{PhysRevLett.114.210801,PhysRevLett.121.060506,Jordan2015}. Another interesting theoretical consideration involves negative quasiprobalities of postselection, demonstrating that postselected experiments can yield anomalously large information-cost rates \cite{Arvidsson-Shukur2020}. 

Since the debates of weak measurement focus on the low postselection efficiency, some targeted efforts have been made on it. It has been demonstrated that by inserting a power-recycling cavity to recycle the failed postselection photons, both the postselection efficiency and signal-to-noise ratio can be significantly improved
\cite{PhysRevLett.114.170801,PhysRevA.88.023821,PhysRevLett.117.230801,FANG2020125117,PhysRevLett.126.220801}. 
The similar conclusion can be achieved by placing a partially transmitting mirror at the dark port of the interferometer, where the improvement of signal scales linearly with the number of cycles \cite{PhysRevA.109.042602}. 
Subsequent theories have proposed that by using a composite dual-recycling cavity, the optimal region of improvement can be further broaden \cite{PhysRevA.108.032608}. 
Another scheme, denoted as joint weak measurement \cite{PhysRevLett.110.083605}, can maximize the utilization rate of photons and remain robust to several sources of noise. 

In recent literature, quantum metrology has been discussed to enhance the precision of weak-value-based measurement beyond the classical methods. By making use of quantum resources such as squeezing and entanglement, there are proposals to break the standard quantum limit (SQL) or even achieve Heisenberg-limited scaling in WVA metrology \cite{PhysRevLett.113.030401,PhysRevA.99.032120,PhysRevLett.115.120401,Starek:20}. In addition, some WVA-based schemes can achieve the Heisenberg limit (HL) without using entanglement \cite{PhysRevLett.128.040503} or even quantum resources \cite{PhysRevLett.114.210801,PhysRevLett.121.060506,Chen2018}, providing a novel perspective for quamtum-enhanced metrology. 

While the introduction of non-classical states indeed enhances the metrological precision of weak measurements, it does not address the issue of low postselection efficiency. Therefore, it is essential to explore the combination of recycling techniques with quantum gain in weak measurements.
In this paper, we try to insert a power-recycling cavity into entanglement-assisted WVA-based scheme of \cite{PhysRevLett.113.030401,PhysRevA.92.012120}. We will show its advantages in improving the performance of WVA-based metrology. Furthermore, we will introduce two types of errors: walk-off errors and readout errors, and analyze the robustness of entanglement-assist cyclic weak measurement against them. The result suggests that entanglement, combined with recycling technique, can further suppress technique noise in weak measurements.

\section{Entanglement-assist weak-value-amplification setup}

We first review the entanglement-assist WVA setup introduced by Pang \textit{et al}. \cite{PhysRevLett.113.030401}. The weak interaction Hamiltonian that couples the system to the pointer variable can be written as

\begin{equation}
    \hat{H}_{int}=\hbar g\hat{A}\otimes \hat{F} \delta \left ( t-t_{0}  \right ),
\end{equation}
where $\hat{A}$ is an observable of the system, $\hat{F}$ is the meter observable, and $g$ is the weak coupling parameter to be estimated. The time factor $\delta \left ( t-t_{0}  \right )$ indicates that the interaction is impulsive. In this system, the meter is coupled to $n$ single-system observables so that $\hat{A}$ is the sum of $n$ observables,
\begin{equation}
    \hat{A} = \hat{A}_1+ \cdots+ \hat{A}_n,
\end{equation}
where $\hat{A}_k= \hat{I}\otimes \cdots \hat{a}  \cdots \otimes \hat{I} $ is shorthand for the observable $\hat{a}$ of the $k$th ancilla. To be more specific, we consider a photonic system and the single-system observable is $\hat{a}=\hat{\sigma_z}$. In this case, the prepared joint state and postselected state are $\left | \psi_i  \right \rangle=\frac{1}{\sqrt{2} } \left ( \left | 0  \right \rangle^{\otimes n} +e^{i\theta }\left | 1 \right \rangle^{\otimes n}     \right )$ and $\left | \psi_f  \right \rangle=\frac{1}{\sqrt{2} } \left ( e^{-in\phi} \left | 0  \right \rangle^{\otimes n} +e^{in(\theta-\pi)}e^{in\phi }\left | 1 \right \rangle^{\otimes n}     \right ) $, respectively, where $e^{i\theta}$ is an arbitrary relative phase and $\phi$ is the parameter corresponding to the postselected angle. We assume $n\phi \ll 1$ and $\theta=0$ for simplicity. Therefore, we can obtain the weak value of $A_w = \langle \psi_f | \hat{A}  | \psi_i   \rangle / \langle \psi_f  | \psi_i   \rangle = -in\cot{n\phi}$ and the postselection probability $P \approx n^2\phi^2$, which is $n$ times that of the case of no entanglement. 
Here, we use the setup introduced by Chen \textit{et al}.\cite{PhysRevA.99.032120}, where the Pauli \textit{Y} operator $\hat{\sigma}_y$ is the meter observable. According to Eqs. (B1) $\sim$ (B4) in  \cite{PhysRevA.99.032120}, the joint state after the weak coupling satisfies
\begin{equation}
    \begin{aligned}
        | \Psi_T   \rangle 
        &= e^{-ig\hat{A}\otimes\hat{\sigma }_y  } | \psi_i   \rangle   | \Phi_0   \rangle \\
        &=[\cos ng-i\frac{\hat{A} }{n}(\sin {ng}) \hat{\sigma}_y  ]| \psi_i   \rangle   | \Phi_0   \rangle,\\
    \end{aligned}   
\end{equation}
where $| \Phi_0   \rangle$ is the initial meter state. 
After the system is postselected into $\left|  \psi_f \right \rangle$, the meter state becomes

\begin{equation}\label{e4}
    \begin{aligned}
        | \Phi_f  \rangle &=  \langle \psi_f  | \Psi_T   \rangle \\
        &=  \langle \psi_f | \psi_i   \rangle \left[ \cos ng-i\frac{A_w}{n}(\sin ng)\hat{\sigma}_y \right]| \Phi_0  \rangle\\
        &=\left[  i\sin n\phi \cos ng - i (\cos n\phi \sin ng)\hat{\sigma}_y   \right]| \Phi_0  \rangle.
    \end{aligned}   
\end{equation}

Next, we deduce the postselection probability for the this case. The state orthogonal to the postselected state is $\left | \psi_f  \right \rangle=\frac{1}{\sqrt{2} } \left ( e^{-in\phi} \left | 0  \right \rangle^{\otimes n} +e^{in\phi }\left | 1 \right \rangle^{\otimes n}     \right ) $. Therefore, the orthogonal meter state can be calculated as 
\begin{equation}
    \begin{aligned}
        | \Phi_f^{\bot }   \rangle =  \langle \psi_f^{\perp }   | \Psi_T   \rangle =\left[  \cos n\phi \cos ng + (\sin n\phi \sin ng)\hat{\sigma}_y   \right]| \Phi_0  \rangle.
    \end{aligned}
\end{equation}
According to the definition of $i\hat{\sigma}_y=| 0_p   \rangle \langle 1_p |-| 1_p   \rangle \langle 0_p |$, we assume that $| \Phi_0   \rangle$ is $| 0   \rangle_p$ and easily obtain 
\begin{equation}
    | \Phi_f  \rangle =  i\sin n\phi \cos ng | 0   \rangle_p+\cos n\phi \sin ng | 1   \rangle_p
\end{equation}
and 
\begin{equation}\label{e7}
    | \Phi_f^{\bot }   \rangle =  \cos n\phi \cos ng |0  \rangle_p+ i(\sin n\phi \sin ng)    |1 \rangle_p.
\end{equation}
Note that we perform measurements on the meter observables $\sigma_R$ using bases $|R\rangle_p$ and $|L\rangle_p$, where $|R\rangle_p=(|0  \rangle_p + i|1  \rangle_p)/\sqrt{2}$ and $|L\rangle_p=(|0  \rangle_p - i|1  \rangle_p)/\sqrt{2}$, so the postselection probability can be calculated as follow:
\begin{equation}
    P=\frac{N}{N+N^{\bot}}=\frac{n^2\cos^2 {ng}+|A_w|^2\sin^2{ng}}{n^2+|A_w|^2}
\end{equation}
where $N=|\langle R_p | \Phi_f\rangle|^2+|\langle L_p | \Phi_f\rangle|^2$ and $N^{\bot}=|\langle R_p | \Phi_f^{\bot}\rangle|^2+|\langle L_p | \Phi_f^{\bot }\rangle|^2$. Similarly, we can obtain the failed postselection probability, 
\begin{equation}
    P^{\bot}=\frac{N^{\bot}}{N+N^{\bot}}=\frac{n^2\sin^2 {ng}+|A_w|^2\cos^2{ng}}{n^2+|A_w|^2}. 
\end{equation}
In the weak value parameter range $ng \ll n\phi \ll 1$, the postselection probability is a small value, $P \approx n^2\phi^2$, while the failed-postselection probability is approximately $1$. Although the Heisenberg-scaling precision can be achieved using only this small successful part of photons\cite{PhysRevLett.113.030401,PhysRevA.92.012120}, a large amount of quantum resources are wasted in the postselection process. Thus, there's still room for further optimization in maximizing the utilization of incident resources in this measurement model.

\begin{figure}[t]
\centering
\includegraphics[trim= 0 0 0 0 ,clip, scale=0.39]{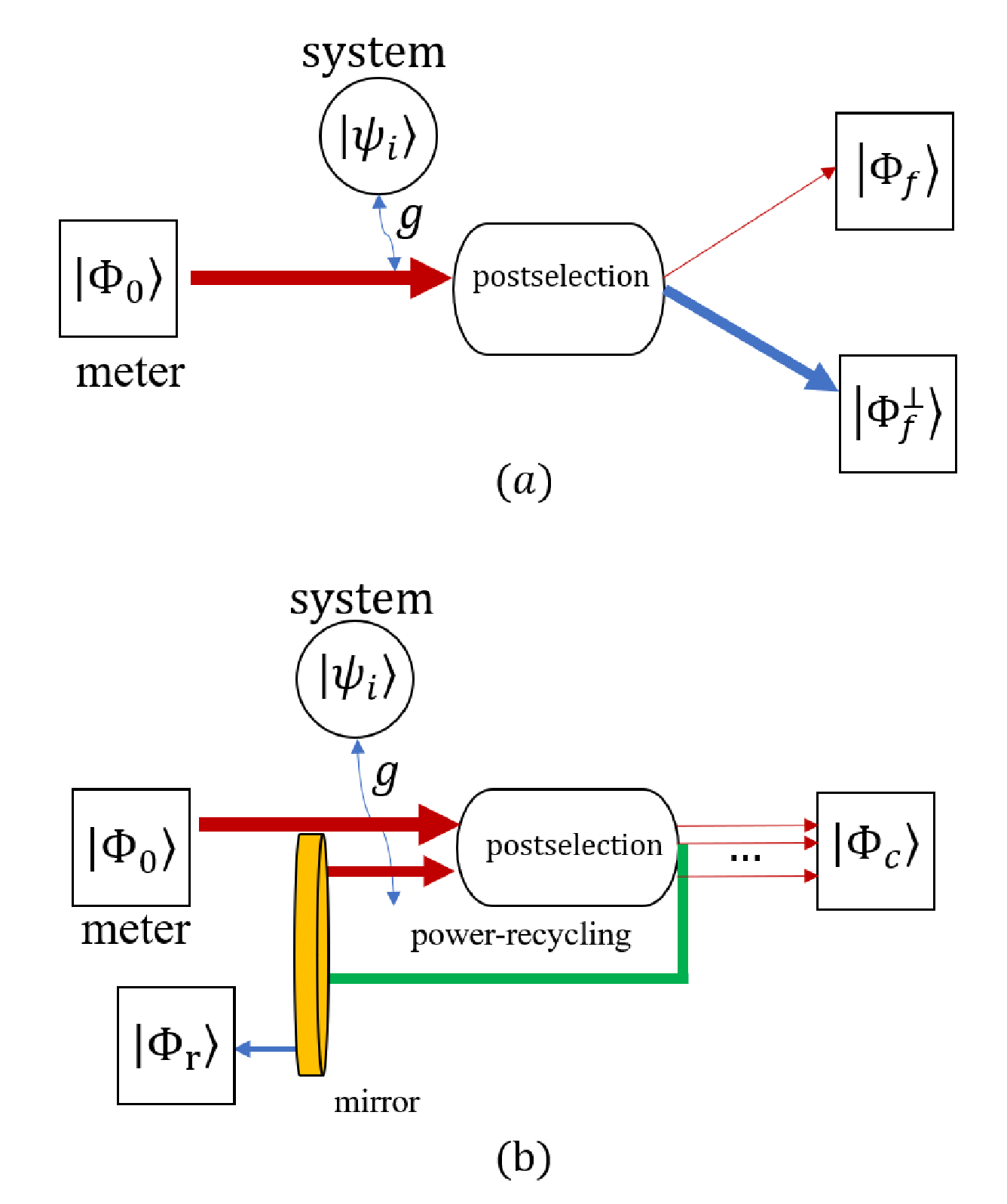}
\caption{(Color online) (a) Sketch of standard WVA-based metrology. The system is initially prepared in state $|\psi_i\rangle$, the incident meter state is $|\Phi_0\rangle$, and the coupling strength between the system and meter is $g$. After postselection, the meter state of successful-postselection part is postselected to $|\Phi_f\rangle$ while the failed-postselection part is $|\Phi_f^{\bot}\rangle$. (b) Sketch of power-recycled WWA-based metrology. Compared to the standard method, the partially transmitting mirror reflects the failed-postselection photons back to the system repeatedly, thereby enhancing the detection efficiency. $|\Phi_c\rangle$: the meter state improved by power-recycling. $|\Phi_r\rangle$: the meter state of unutilized photons. } 
\label{Fig.1}
\end{figure}

\section{Power-recycling}
\begin{figure}[t]
\centering
\includegraphics[trim= 0.1 0.1 0.1 0.1 ,clip, scale=0.5]{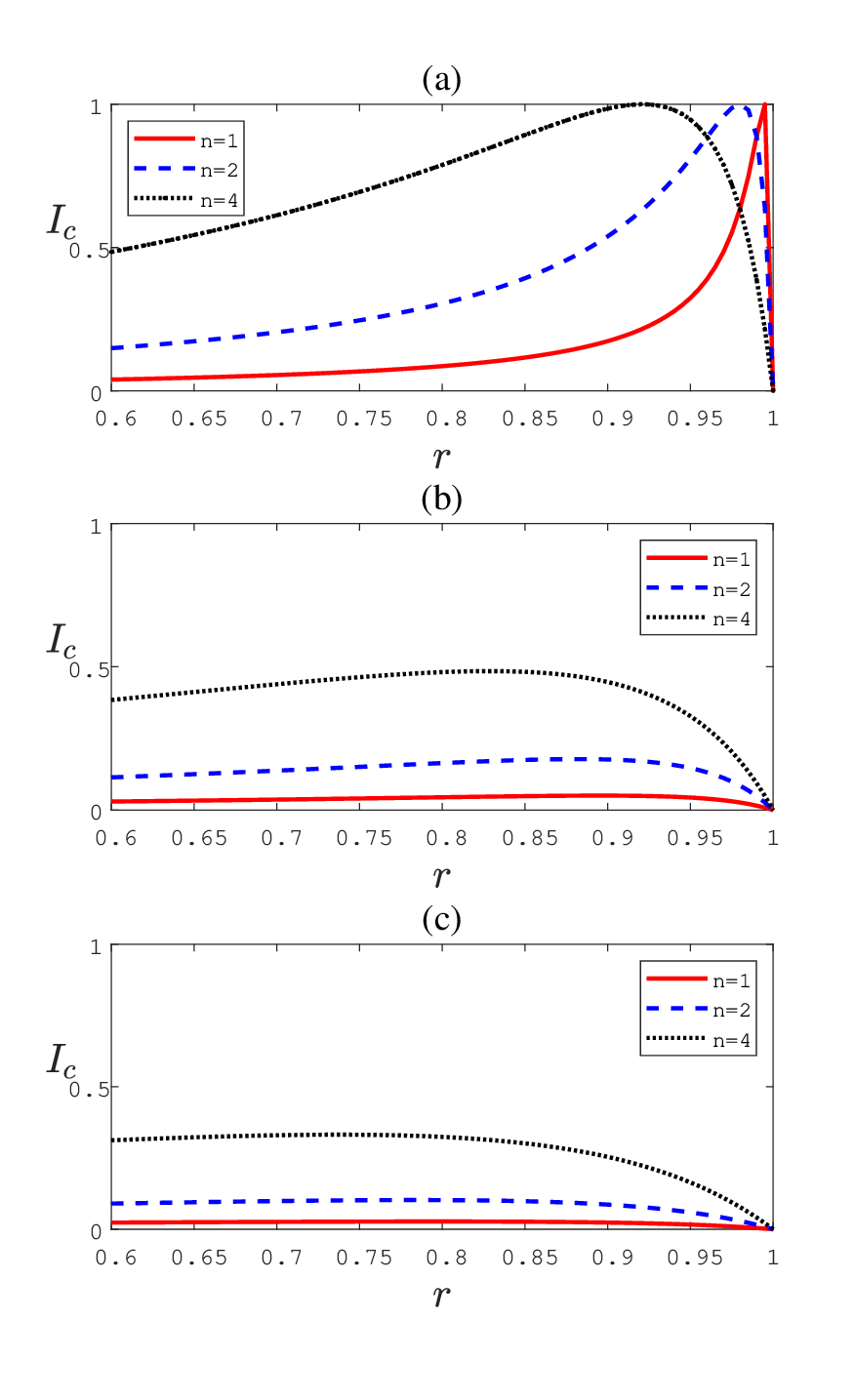}
\caption{(Color online) Panels (a), (b), and (c) correspond to the cases of $I_c$ varying with $r$ under different values of $\gamma$: 0, 0.05, and 0.1, where $\phi=0.1$. The red solid, blue dashed and black dotted lines correspond to the entangled qubits $n=1,2,4$, respectively. $I_c$ is the detected intensity of power-recycled entanglement-assisted WWA scheme, $r$ is the reflection coefficient of partially transmitting mirror, $\gamma$ is the optical loss, and $\phi$ is the parameter corresponding to the postselected angle. }
\label{Fig.2}
\end{figure}

We consider reusing the failed-postselection photons via power-recycling, as shown in Fig. \ref{Fig.1}(b). A partially transmitting mirror forms an optical cavity in combination with the measurement system, reusing the failed-postselection signals\cite{PhysRevLett.114.170801,PhysRevA.88.023821}. We can approximate that the cyclic process won't disrupt the entangled system states since there's no measurement involved. Similar to the discussion in \cite{PhysRevLett.114.170801,FANG2020125117,PhysRevA.108.032608,PhysRevA.109.042602}, we use $\hat{U}_g=e^{-ig\hat{A}\hat{\sigma }_y  }$ to express the weak interaction and $\hat{U}_{\phi}=e^{-i\hat{A}\phi}$ to represent the postselection process. We also introduce two orthogonal system states $\left | \psi_+  \right \rangle=\frac{1}{\sqrt{2} } \left ( \left | 0  \right \rangle^{\otimes n} +\left | 1 \right \rangle^{\otimes n}     \right )$ and $\left | \psi_-  \right \rangle=\frac{1}{\sqrt{2} } \left ( \left | 0  \right \rangle^{\otimes n} -\left | 1 \right \rangle^{\otimes n}     \right )$, which have the relationship with initial and postselected states, $\left | \psi_i  \right \rangle=\left | \psi_+  \right \rangle$ and $\left | \psi_f  \right \rangle=\hat{U}_{\phi}\left | \psi_-  \right \rangle$. Thus, the meter state postselected by the output ends is 
\begin{equation}
    \begin{aligned}
        | \Phi_f  \rangle &=  \langle \psi_f  | \Psi_T   \rangle=\langle \psi_- | \hat{U}_{\phi}^{\dagger }  \hat{U}_g | \psi_+ \rangle | \Phi_0  \rangle \\
        &=  i\sin {(n\phi-ng\hat{\sigma}_y)}| \Phi_0  \rangle,\\
    \end{aligned}   
\end{equation}
which is equivalent to Eq. (\ref{e4}). Similarly, the meter state postselected by the input end is 
\begin{equation}
    \begin{aligned}
        | \Phi_f^{\bot}  \rangle &=  \langle \psi_f^{\bot}  | \Psi_T   \rangle=\langle \psi_+ | \hat{U}_{\phi}^{\dagger }  \hat{U}_g | \psi_+ \rangle | \Phi_0  \rangle \\
        &=  \cos {(n\phi-ng\hat{\sigma}_y)}| \Phi_0  \rangle.\\
    \end{aligned}   
\end{equation}
Next, we assume that the cavity is resonant and denote mirror's reflection and transmission coefficient by $r$ and $p$ ($r^2+p^2=1$), respectively. Considering the imperfection of the optical system, we introduce the non-unitary operator $\hat{L} =\sqrt{1-\gamma}$, where $\gamma$ is the single-pass power loss, to express the loss of optical imperfection in one return. Therefore, the meter state improved by power-recycling is given by the sum of successful-postselection meter states from all traversal numbers,
\begin{equation}\label{e12}
    \begin{aligned}
        | \Phi_c  \rangle &=  p\langle \psi_- | \hat{U}_{\phi}^{\dagger }  \hat{U}_g | \psi_+ \rangle \sum_{n=0}^{\infty}  \left(r\hat{L}\langle \psi_+ | \hat{U}_{\phi}^{\dagger }  \hat{U}_g | \psi_+ \rangle \right)^n| \Phi_0  \rangle\\
        &\approx  \frac{ip\sin {(n\phi-ng\hat{\sigma}_y)}}{1-r\hat{L}\cos {(n\phi-ng\hat{\sigma}_y)}}| \Phi_0  \rangle,\\
    \end{aligned}   
\end{equation}
where the approximate is taken in the condition of $( n\hat{L})^{\infty}=0$. Taking into consideration that $ng \ll n\phi \ll 1$ and $| \Phi_0 \rangle=| 0 \rangle_p$, Eq. (\ref{e12}) can be further simplified via Taylor expansion:
\begin{equation}\label{e13}
    \begin{aligned}
        &| \Phi_c  \rangle \approx \frac{ip\sin {n\phi}}{1-r\hat{L}\cos {n\phi}} \left[  1 - \frac{ng\hat{\sigma}_y(\cos {n\phi}-r\hat{L})}{\sin {n\phi}(1-r\hat{L}\cos {n\phi})} \right]| 0 \rangle_p\\
        &=\frac{ip\sin {n\phi}}{1-r\hat{L}\cos {n\phi}}\left[  | 0 \rangle_p - ing\frac{\cos {n\phi}-r\hat{L}}{\sin {n\phi}(1-r\hat{L}\cos {n\phi})} | 1 \rangle_p  \right],\\
    \end{aligned}   
\end{equation}
where we neglect high-order terms of $ng$. The power of detected light is then given by 
\begin{equation}\label{e14}
    \begin{aligned}
        I_c=|\langle \Phi_c | \Phi_c  \rangle|^2 \approx \frac{p^2\sin ^2{n\phi}}{(1-r\hat{L}\cos n\phi)^2}.
    \end{aligned}   
\end{equation}
Since the incident light is normalized, $I_c$ can denote the fraction of incident light being detected. As shown in Fig. \ref{Fig.2}, we assume $\phi=0.1$ and plot $I_c$ varying with $r$ under different conditions, where $\gamma=0, 0.05, 0.1$ and $n=1,2,4$. Our findings indicate that: (i) entanglement cannot enhance the theoretical peak power. (ii) higher entangled ancillas and lower optical loss can improve the recovery rate of photons. 

\section{Fisher information analysis}

\begin{figure}[t]
\centering
\includegraphics[trim= 0.1 0.1 0.1 0.1 ,clip, scale=0.5]{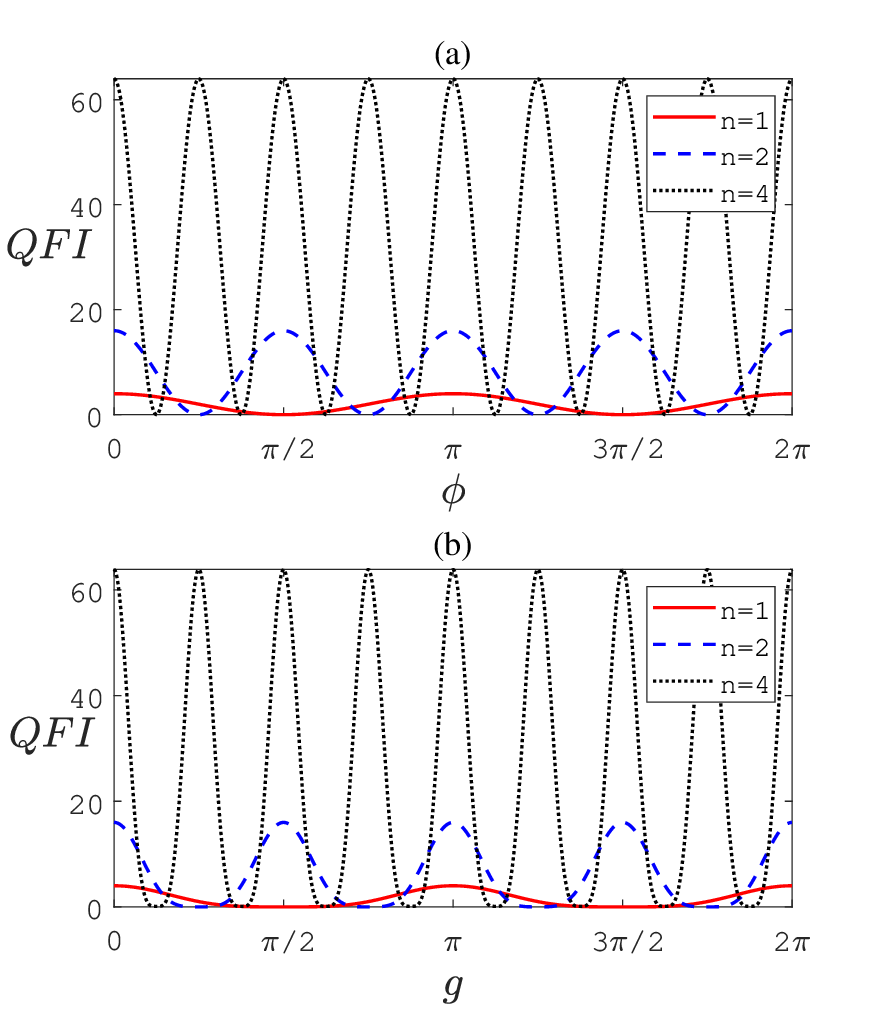}
\caption{(Color online) The quantum Fisher information (QFI) of standard entanglement-assisted protocol. Panels (a) and (b) correspond to the cases of QFI varying with $\phi$ and $g$, respectively. In these plots, the red solid, blue dashed, and black dotted lines represent different entangled qubits, denoted as $n=1,2,4$. $\phi$ is the parameter corresponding to the postselected angle and $g$ is the coupling strength between the system and meter. }
\label{Fig.3}
\end{figure}

The Fisher information (FI) is a central quantity in parameter estimation, denoted by $\mathcal{I}(g)$ in this article. It can low-bound the variance of an unbiased estimator $g$ via the Cramér-Rao inequality: $Var(g)\ge 1/\mathcal{I}(g)$ \cite{Wiseman_Milburn_2009}. The quantum Fisher information (QFI) is defined as the FI maximized over all possible generalized measurements. In the case where the meter function $|\Phi\rangle$ is known, the QFI can be calculated using the formula
\begin{equation}\label{e15}
    \mathcal{QI}(g)= 4\left[(\frac{d\left \langle \Phi\right | }{dg } )(\frac{d\left | \Phi  \right \rangle }{dg } ) - {\left |\left \langle \Phi \right| (\frac{d\left | \Phi  \right \rangle }{dg } ) \right |}^2 \right].
\end{equation}
By substituting (\ref{e4}) into (\ref{e15}), we can obtain the QFI of $| \Phi_f \rangle$
\begin{equation}
    \begin{aligned}
        \mathcal{QI}_f^{(n)} & = 4n^2\left( \sin^2 {ng} \cos^2 {n\phi} + \cos^2 {ng} \sin^2 {n\phi}\right) \\
        &- n^2(\sin^2 {2ng} \cos^2 {2n\phi})^2.
    \end{aligned} 
\end{equation}
In Fig. \ref{Fig.3}, we plot $\mathcal{QI}_f^{(n)}$ varying with $\phi$ (Fig. \ref{Fig.3}(a)) and $g$ (Fig. \ref{Fig.3}(b)) under the entangled ancillas $n=1,2,4$, respectively. It shows that the Heisenberg-scaling, $\mathcal{QI}\approx 4n^2$, can be achieved if providing proper weak-coupling and postselection parameters range. Especially, the weak value range ($ng\ll n\phi\ll1$) is included, which implies a significant advantage of weak measurements: achieving high-precision estimation with minimal photon resources. 

\begin{figure}[t]
\centering
\includegraphics[trim= 0.1 0.1 0.1 0.1 ,clip, scale=0.5]{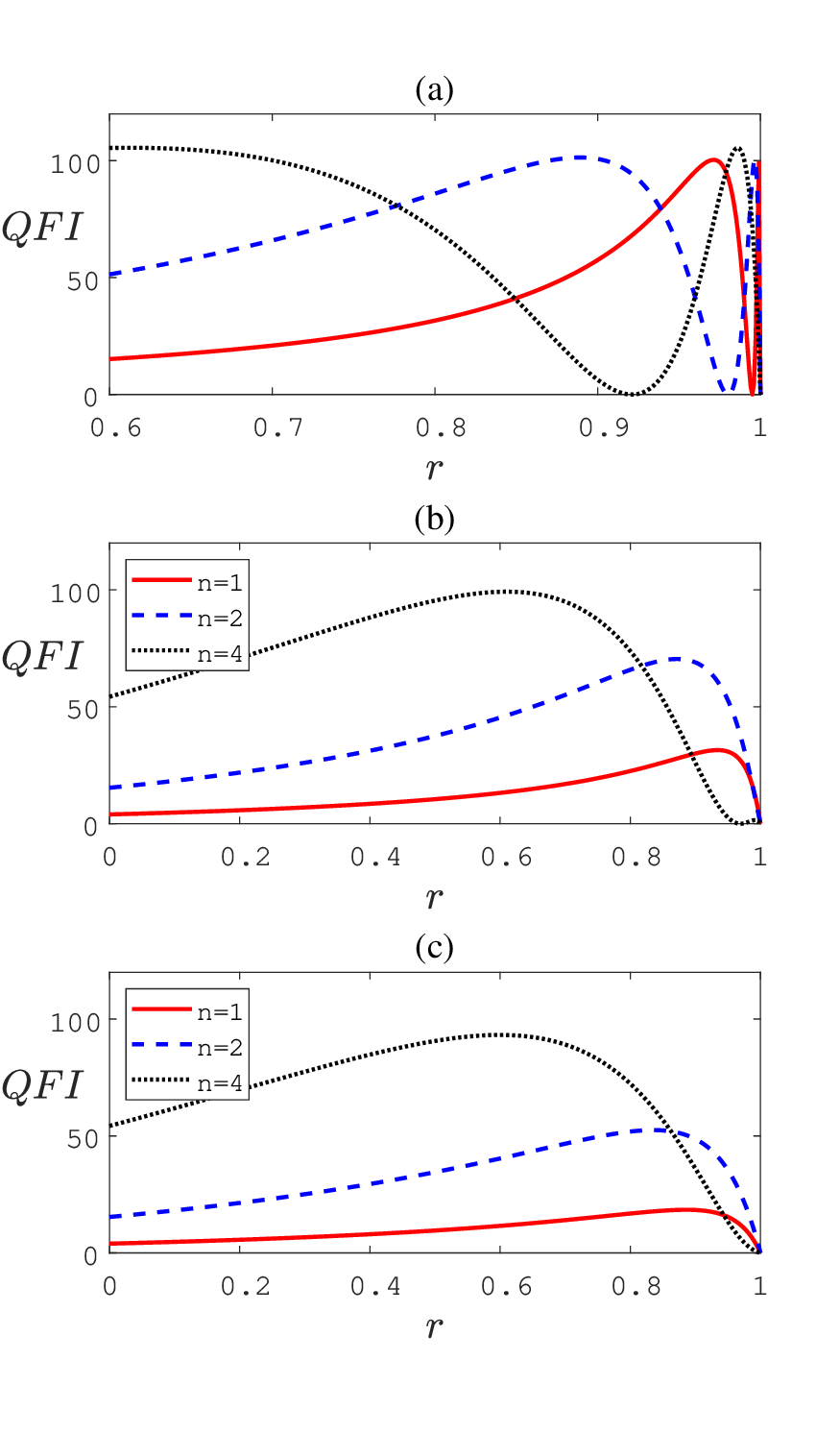}
\caption{(Color online) The quantum Fisher information (QFI) of power-recycled entanglement-assisted protocol. Panels (a), (b), and (c) correspond to the cases of $\mathcal{QI}_c^{(n)}$ varying with $r$ under different values of $\gamma$: 0, 0.05, and 0.1, where $\phi=0.1$. The red solid, blue dashed and black dotted lines correspond to the entangled qubits $n=1,2,4$, respectively. $\mathcal{QI}_c^{(n)}$ is the QFI of power-recycled entanglement-assisted WWA scheme, $r$ is the reflection coefficient of partially transmitting mirror, $\gamma$ is the optical loss, and $\phi$ is the parameter corresponding to the postselected angle.}
\label{Fig.4}
\end{figure}

Similarly, we can obtain the QFI of $|\Phi_c\rangle$ by substituting (\ref{e13}) into (\ref{e15}):
\begin{equation}
    \begin{aligned}
        \mathcal{QI}_c^{(n)} & \approx 4n^2p^2\left[ \frac{(\cos {n\phi}-r\hat{L})^2}{(1-r\hat{L}\cos {n\phi})^4} - g^2\left| \frac{\cos {n\phi}-r\hat{L}}{1-r\hat{L}\cos {n\phi}}\right|^2 \right]\\
        &\approx 4n^2\frac{p^2(\cos {n\phi}-r\hat{L})^2}{(1-r\hat{L}\cos {n\phi})^4},
    \end{aligned} 
\end{equation}
where we neglect high-order terms of $ng$. It can be seen that the QFI of power-recycling is boosted by a factor $p^2(\cos {n\phi}-r\hat{L})^2/(1-r\hat{L}\cos {n\phi})^4$ compared to the conventional scheme or single-pass weak-value scheme. As shown Fig. \ref{Fig.4}, we set $\phi=0.1$ and plot $\mathcal{QI}_c^{(n)}$ varying with $r$ under different entangled ancillas $n$ and optical loss $\gamma$. (i) as expected, the power-recycling can really further improve the QFI of detection. However, we have to declare that it does not means that the Heisenberg limit is exceeded since the improvement originates from the increasing of detected photons, without changing the rule $\mathcal{QI}\approx 4n^2$ \cite{PhysRevA.109.042602}. (ii) higher entangled ancillas can provide wider optimal region of precision. (iii) As $n$ increases, the impact of $\gamma$ on the QFI decreases. This implies that even in real experiments with high optical losses, a high level of precision can be maintained,  which is well-suited for experimental research. 

\begin{figure}[t]
\centering
\includegraphics[trim= 0.1 0.1 0.1 0.1 ,clip, scale=0.54]{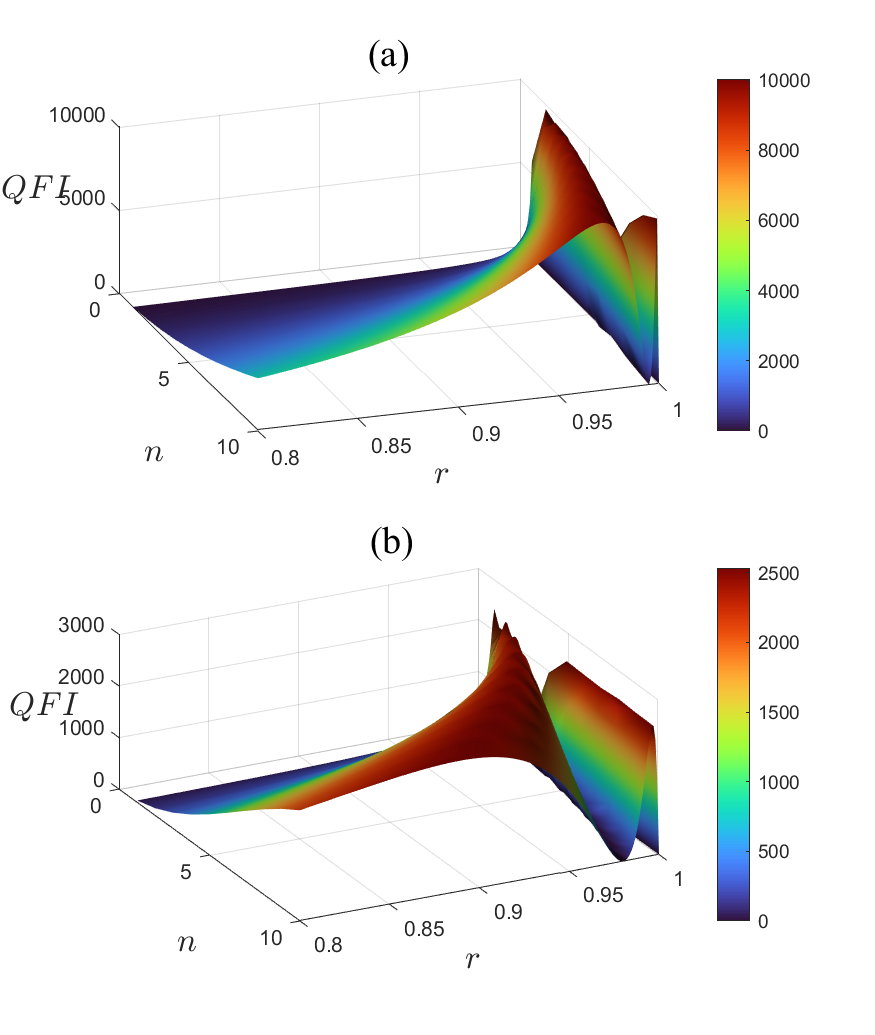}
\caption{(Color online) $\mathcal{QI}_c^{(n)}$ in 3D. Panels (a) and (b) correspond to two different postselection distributions $\phi=0.01$ and $\phi=0.02$, respectively, under ideal optical loss. The small value of $\phi$ is chosen to satisfy the weak value region of $n\phi \ll 1$. These two figures indicate that the power-recycling component determines the upper bound of $\mathcal{QI}_c^{(n)}$. $\mathcal{QI}_c^{(n)}$ is the quantum Fisher information of power-recycled entanglement-assisted WWA scheme, $r$ is the reflection coefficient of partially transmitting mirror, $\phi$ is the parameter corresponding to the postselected angle, and $n$ is the number of entangled qubits. }
\label{Fig.5}
\end{figure}

To visually represent the variation of $\mathcal{QI}_c^{(n)}$ with respect to the entangled ancillas $n$, we set a small value for $\phi$ to expand the range of $n$ (since $n\phi \ll 1$), and plot a 3-dimensional graph in Fig. \ref{Fig.5}. We find that the peak value of $\mathcal{QI}_c^{(n)}$ depends on the parameters of the resonant cavity but is independent of the entanglement. This conclusion is reasonable since within the weak value regime, the precision limit of a $n$-uncorrected power-recycled weak measurement systems satisfy $n\mathcal{QI}_c^{(1)} \sim  n/\phi$ \cite{PhysRevLett.114.170801,PhysRevA.88.023821,PhysRevX.4.011031}, which surpasses $\mathcal{QI}_f \sim n^2$ of the $n$-entangled weak measurement system. Therefore, in the weak measurement system involving entanglement and power-recycling, the recycling component determines the upper limit of precision, while the entanglement component broadly improves the performance under various parameters.

\section{Error analysis}
In previous sections, we provided a basic model of entanglement-enhanced cyclic weak measurement. However, in practical implementations, there are some issues should be taken into consideration, especially the stability of the resonator, signal offsets and readout errors. 

Generally, the length of the resonance cavity is unstable and time dependent due to the influence of platform jitter, temperature, pressure and so on. Wang \textit{et al}. proposed using the Pound-Drever-Hall (PDH) technique \cite{Drever1983} for the cavity locking and successfully applied it in power-recycling weak measurement \cite{PhysRevLett.117.230801}. Indeed, the utilization of differential signals for feedback regulation in this technique remains unaffected by the influence of entanglement, rendering it equally applicable to our system. 

In \cite{PhysRevA.88.023821}, Dressel \textit{et al}. proposed that the recycling beam tends to diminish the signal after many traversals, a phenomenon known as the walk-off effect. This phenomenon constitutes the primary rationale behind the dissonance observed between Fisher information gain and the power gain of the detected signal. In previous models utilizing Gaussian beam profiles, the walk-off effect could be mitigated using filtering principles \cite{PhysRevLett.114.170801,PhysRevA.109.042602}. However, for the entanglement-assist model, we have yet to find a method to completely eliminate the walk-off effect. Therefore, in Section \ref{A}, we will significantly discuss the impact of the walk-off effect on the measurement result. 

Readout errors, which result from noise in the environment and technical imperfections in the experimental devices, can significantly influence the weak measurement protocol. Based on the readout error model constructed in literature \cite{PhysRevA.92.012120}, we  will further investigate the suppressive effect of power-recycling on readout errors in Section \ref{B}. 

\begin{figure}[t]
\centering
\includegraphics[trim= 0.1 0.1 0.1 0.1 ,clip, scale=0.54]{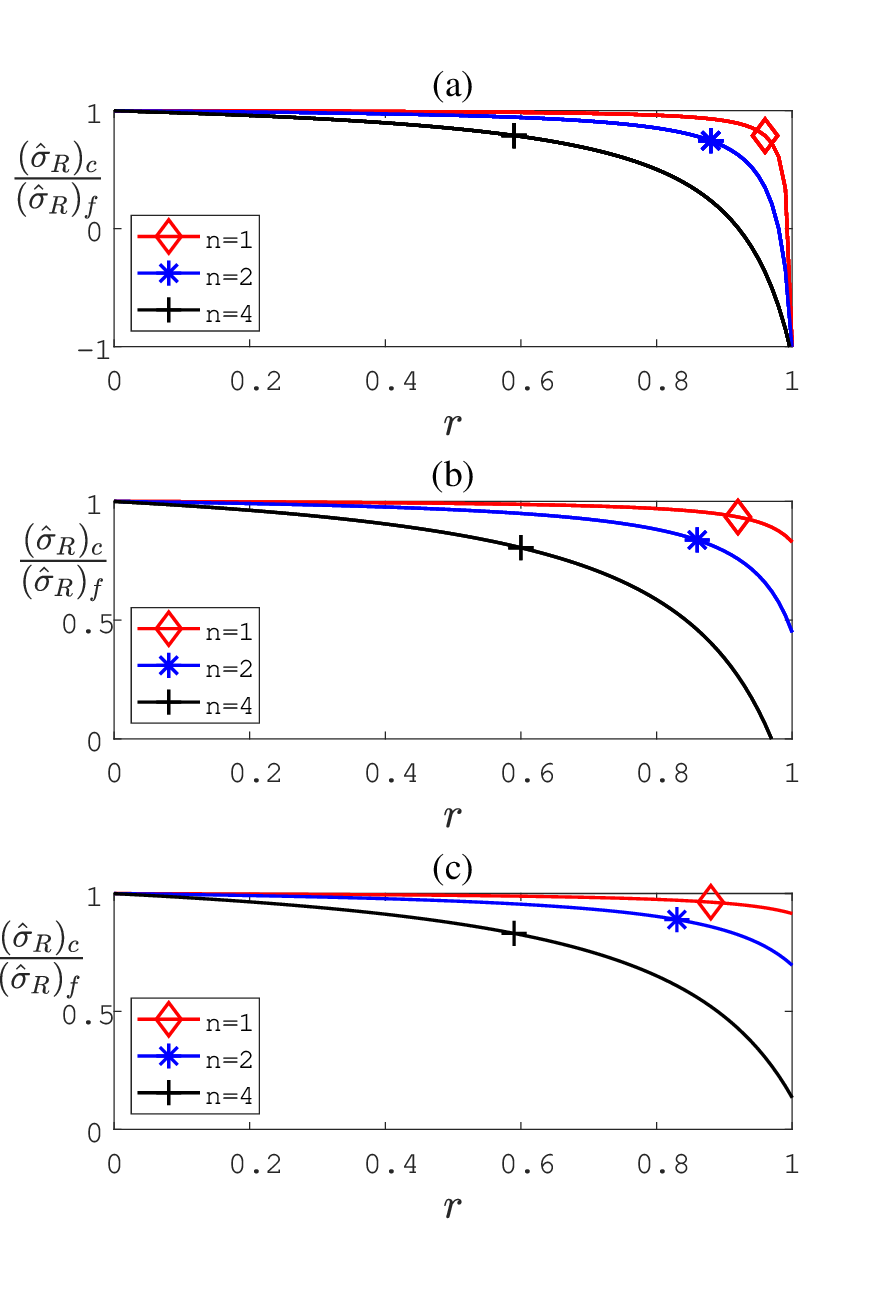}
\caption{(Color online) The walk-off effect induced by power-recycling. Panels (a), (b), and (c) correspond to the cases of $\left \langle \hat{\sigma } _R \right \rangle_c/\left \langle \hat{\sigma } _R \right \rangle_f$ varying with $r$ under different values of $\gamma$: 0, 0.05, and 0.1, where $\phi=0.1$. The red solid, blue dashed and black dotted lines correspond to the entangled qubits $n=1,2,4$, respectively. $\left \langle \hat{\sigma } _R \right \rangle_c/\left \langle \hat{\sigma } _R \right \rangle_f$ is the ratio of measurement results obtained using the power-recycling method to those obtained using the standard method, $r$ is the reflection coefficient of partially transmitting mirror, $\gamma$ is the optical loss, and $\phi$ is the parameter corresponding to the postselected angle.}
\label{Fig.6}
\end{figure}

\subsection{Walk-off errors}\label{A}
In this section, we analyze the effect of walk-off error on the measurement result. Since the weak value here is imaginary, we choose $\hat{\sigma}_R$, which satisfies $\hat{\sigma}_R=\begin{pmatrix}
  0&-i \\
  i&0
\end{pmatrix}$ in our representation, as the pointer observable. We first consider the scenario without power-recycling. In standard entanglement-assisted weak-value-amplification scheme, the meter state is $|\Phi_f\rangle$. So the expectation value of $\hat{\sigma}_R$ is
\begin{equation}\label{e18}
    \begin{aligned}
        &\left \langle \hat{\sigma } _R \right \rangle_f =\frac{\left \langle \Phi_f  |\hat{\sigma } _R |\Phi_f  \right \rangle }{\left \langle \Phi_f  |\Phi_f  \right \rangle} \\
        &=-\frac{2\sin n\phi \cos n\phi \sin ng \cos ng}{\sin^2 n\phi \cos^2 ng + \cos^2 n\phi \sin^2 ng}.
    \end{aligned} 
\end{equation}
With the power-recycling cavity, the meter state becomes $| \Phi_c  \rangle$. By the same method, the expectation value of $\hat{\sigma}_R$ changes to 
\begin{equation}\label{e19}
    \begin{aligned}
        &\left \langle \hat{\sigma } _R \right \rangle_c = \frac{\left \langle \Phi_c  |\hat{\sigma } _R |\Phi_c  \right \rangle }{\left \langle \Phi_c |\Phi_c  \right \rangle} \\
        &\approx -\frac{2ng\frac{\cos n\phi - r\hat{L}}{(1-r\hat{L}\cos n\phi)\sin n\phi}}{1+(ng)^2\left[  \frac{\cos n\phi - r\hat{L}}{(1-r\hat{L}\cos n\phi)\sin n\phi} \right]^2}\\
        &\approx -\frac{2ng(\cos n\phi - r\hat{L})}{(1-r\hat{L}\cos n\phi)\sin n\phi}, 
    \end{aligned} 
\end{equation}
where we neglect high-order terms of $ng$. In Fig. \ref{Fig.6}, we plot $\left \langle \hat{\sigma } _R \right \rangle_c/\left \langle \hat{\sigma } _R \right \rangle_f$ varying with $r$ under different conditions. First, the power-recycling reduces the measurement result, which aligns with the conclusions drawn from typical weak measurement systems employing Gaussian pointer. This is because the failed-postselection photons utilized in the recycling carry partial measurement information, which tends to counteract the offset introduced by weak coupling. Moreover, under certain conditions, this partial information can completely nullify the offset (For instance, at the point where Fisher information drops to zero as depicted in Fig. \ref{Fig.4}(a), there are photons detected yet no Fisher information is present). Second, entanglement increases the walk-off error. Under the same circumstances, a higher entangled ancillas corresponds to a lower ratio of $\left \langle \hat{\sigma } _R \right \rangle_c$ to $\left \langle \hat{\sigma } _R \right \rangle_f$. Third, entanglement induces a movement of the optimal region of precision towards smaller walk-off errors. As shown in Fig. \ref{Fig.6}, we mark the walk-off error corresponding to maximum QFI and observe that the marker moves almost parallel to the left as $n$ increases, which tends to counteract walk-off errors induced by the entanglement. Therefore, a higher entangled qubit number $n$ enable the attainment of greater Fisher information under equivalent walk-off error conditions.

\begin{figure*}[t]
\centering
\includegraphics[trim= 0.1 0.1 0.1 0.1 ,clip, scale=0.65]{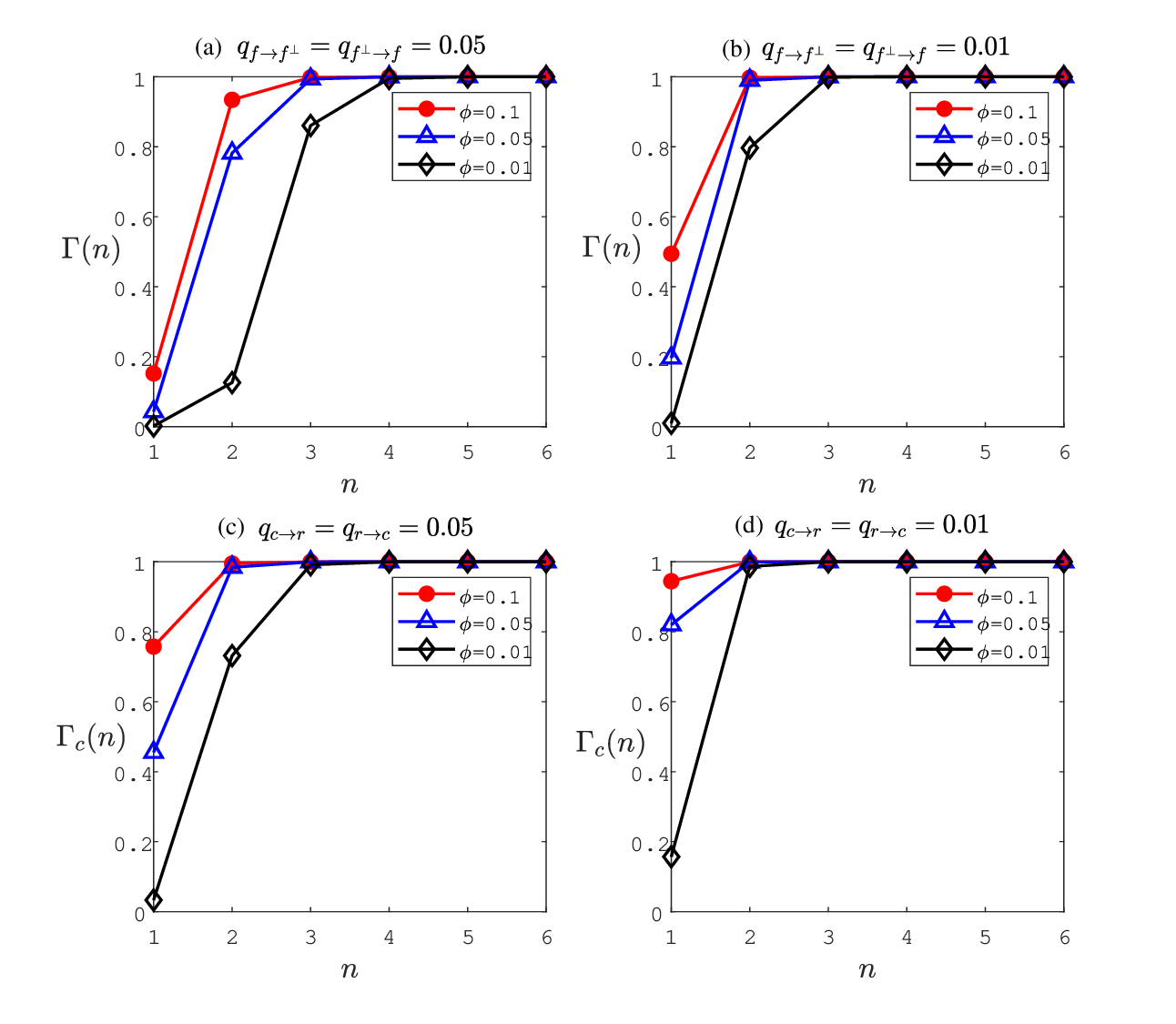}
\caption{(Color online) Comparison of the effect of readout errors on the measurement results. (a) and (b) plot $\Gamma(n)$ versus the number of entangled qubits $n$ for different values of $\phi$, whereas (c) and (d) show the trends of $\Gamma_c(n)$. $\Gamma(n)$, $\Gamma_c(n)$ are the modification factors of the measurement results of standard and power-recycled protocols, respectively. }
\label{Fig.7}
\end{figure*}

\subsection{Readout errors}\label{B}
\subsubsection{Readout errors in standard entanglement-assisted protocol}
We first review the noise model of \cite{PhysRevA.92.012120}, where readout errors significantly influence the weak measurement protocol by distorting the postselected results. Suppose the probability of mistaking $| \Phi_f \rangle$ to $| \Phi_f^{\bot} \rangle$ is $q_{f \to f^{\bot}}$ and the probability of mistaking $| \Phi_f^{\bot} \rangle$ to$| \Phi_f \rangle$ is $q_{ f^{\bot} \to f}$. According to Eq. (86) in \cite{PhysRevA.92.012120}, the postselection probability in entanglement-assisted protocol should be modified as 
\begin{equation}\label{e20}
    \begin{aligned}
        P_{mod}=P(1-q_{f \to f^{\bot}})^n+P^{\bot}q^n_{ f^{\bot} \to f}.
    \end{aligned} 
\end{equation}
$(1-q_{f \to f^{\bot}})^n$ quantifies the loss of correct postselection results, the corresponding loss rate is therefore
\begin{equation}
    \begin{aligned}
       R_{loss}=1-(1-q_{f \to f^{\bot}})^n \approx n q_{f \to f^{\bot}}. 
    \end{aligned} 
\end{equation}
The relative error rate in the postselection results is 
\begin{equation}
    \begin{aligned}
        P_{error}=\frac{P^{\bot}q^n_{ f^{\bot} \to f}}{P(1-q_{f \to f^{\bot}})^n+P^{\bot}q^n_{ f^{\bot} \to f}}.
    \end{aligned} 
\end{equation}

As shown in Eq. (\ref{e18}), we have calculated the expectation value of $\hat{\sigma}_R$ under the pointer state $|\Phi_f\rangle$. According to Eq. (\ref{e7}), the average value of pointer under the pointer state  $|\Phi_f^{\bot}\rangle$ is 
\begin{equation}
    \begin{aligned}
        &\left \langle \hat{\sigma } _R \right \rangle^{\bot}_f =\frac{\left \langle \Phi_f^{\bot}  |\hat{\sigma } _R |\Phi_f^{\bot}  \right \rangle }{\left \langle \Phi_f^{\bot}  |\Phi_f^{\bot}  \right \rangle} \\
        &=\frac{2\sin n\phi \cos n\phi \sin ng \cos ng}{\cos^2 n\phi \cos^2 ng + \sin^2 n\phi \sin^2 ng}.
    \end{aligned} 
\end{equation}
Thus, taking readout errors into account, the measurement result is given by
\begin{equation}
    \begin{aligned}
        \left \langle \hat{\sigma } _R \right \rangle_m &= \frac{P(1-q_{f \to f^{\bot}})^n\left \langle \hat{\sigma } _R \right \rangle_f+P^{\bot}q^n_{ f^{\bot} \to f}\left \langle \hat{\sigma } _R \right \rangle^{\bot}_f}{P(1-q_{f \to f^{\bot}})^n+P^{\bot}q^n_{ f^{\bot} \to f}} \\
        &=\Gamma(n)\left \langle \hat{\sigma } _R \right \rangle_f, 
    \end{aligned} 
\end{equation}
where
\begin{equation}\label{e24}
    \Gamma(n)=\frac{P\left[(1-q_{f \to f^{\bot}})^n-  q^n_{ f^{\bot} \to f} \right]}{P(1-q_{f \to f^{\bot}})^n+P^{\bot}q^n_{ f^{\bot} \to f}}. 
\end{equation}
As readout errors decreases, $\Gamma(n)$ approaches $1$. Conversely, when $\Gamma(n)=0$, readout errors completely destroy the measurement outcome. 

\begin{figure*}[t]
\centering
\includegraphics[trim= 0.1 0.1 0.1 0.1 ,clip, scale=0.6]{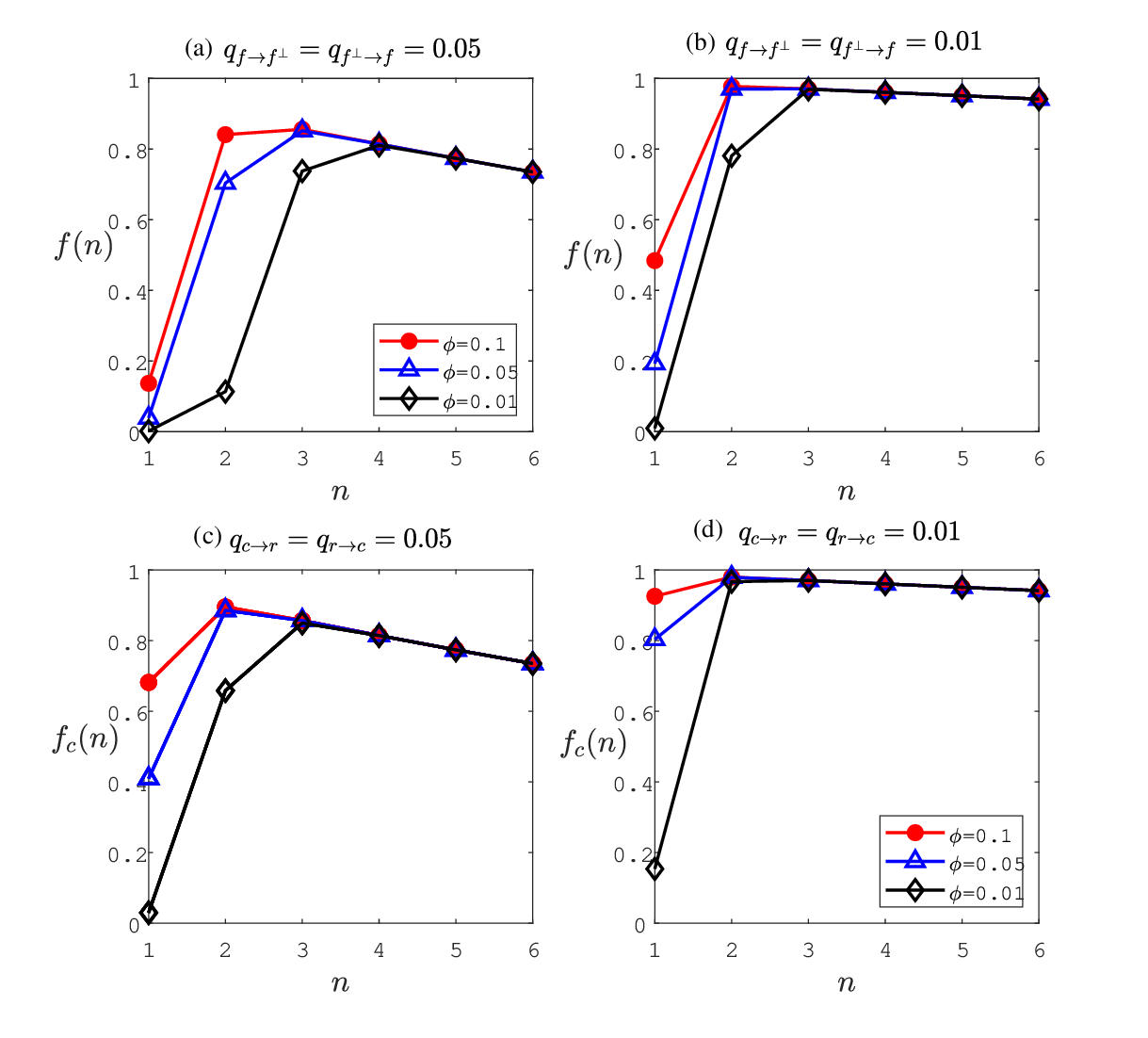}
\caption{(Color online) Comparison of the effect of readout errors on the Fisher information. (a) and (b) plot $f(n)$ varying the number of entangled qubits $n$ for different values of $\phi$, whereas (c) and (d) show the trends of $f_c(n)$. $f(n)$, $f_c(n)$ are the modification factors of the Fisher information of standard and power-recycled protocols, respectively. These figures illustrate that the power-recycling component substantially improves the situation where a significant loss of Fisher information occurs in low-qubit scenarios, while also increasing the peak values of the curves under certain circumstances. }
\label{Fig.8}
\end{figure*}

Next, we introduce the influence of readout errors on the Fisher information of estimating $g$. In Eqs. (102)-(114) of Ref. \cite{PhysRevA.92.012120}, the authors have demonstrated that the Fisher information modified by the readout errors is decreased by a factor $f(n)$, 
\begin{equation}
    \mathcal{I}(g)_{mod}=f(n)\mathcal{I}(g), 
\end{equation}
where 
\begin{equation}
    f(n)=\frac{n^2\left[ (1-q_{f \to f^{\bot}})^n - q_{f^{\bot} \to f}^n \right]^2}{n^2(1-q_{f \to f^{\bot}})^n+|A_w|^2q_{f^{\bot} \to f}^n }. 
\end{equation}
In figures. \ref{Fig.7}(a)(b) and \ref{Fig.8}(a)(b), we plot factors $\Gamma (n)$ and $f(n)$ varying with $n$, respectively. It shows one of the main conclusions of \cite{PhysRevA.92.012120}: entanglement can significantly strengthen the postselected weak measurement against readout errors. In addition, as $n$ further increases, the Fisher information falls since the dramatically increase of loss rate gradually disrupts the postselection. 

\subsubsection{Readout errors in power-recycled entanglement-assisted protocol}

With the power-recycling cavity, most of the failed-postselection photons are reflected back to the weak measurement system and the truly discarded photons are the part reflected back toward the laser. In this way, the meter state of discarding part is  
\begin{equation}\label{e25}
    \begin{aligned}
        &| \Phi_r  \rangle =  \left[  -r+ \frac{p^2\hat{L}\cos {(n\phi-ng\hat{\sigma}_y)}}{1-r\hat{L}\cos {(n\phi-ng\hat{\sigma}_y)}}    \right]| \Phi_0  \rangle\\
        &\approx \frac{\hat{L}\cos {n\phi} -r}{1-r\hat{L}\cos n\phi} |0\rangle_p + ing\frac{p^2\hat{L}\sin n\phi}{(1-r\hat{L}\cos n\phi)^2}|1\rangle_p   ,\\
    \end{aligned}   
\end{equation}
where we use Taylor expansion and neglect high-order terms of $ng$. Similarly, we take $|R\rangle_p$ and $|L\rangle_p$ as bases and perform measurements on the meter observables $\sigma_R$. The detection probability and discarding probability using power-recycling are given by $P_c$ and $P_r$, respectively.
\begin{equation}
    P_c=\frac{N_c}{N_c+N_r}=\frac{p^2 \sin^2 {n\phi}}{p^2 \sin^2 {n\phi}+(\hat{L}\cos n\phi -r)^2}
\end{equation}
and 
\begin{equation}
    P_r=\frac{N_r}{N_c+N_r}=\frac{(\hat{L}\cos n\phi -r)^2}{p^2 \sin^2 {n\phi}+(\hat{L}\cos n\phi -r)^2}, 
\end{equation}
where $N_c=|\langle R_p | \Phi_c\rangle|^2+|\langle L_p | \Phi_c\rangle|^2$ and $N_r=|\langle R_p | \Phi_r\rangle|^2+|\langle L_p | \Phi_r\rangle|^2$. It can be seen that power recycling directly modifies the distribution of postselection, which should lead to changes in readout errors that are sensitive to the postselection. Next, we will demonstrate the suppressive effect of power recycling on readout errors.

Similarly, we assume the probability of mistaking $|\Phi_c\rangle$ to $|\Phi_r\rangle$ is $q_{c \to r}$ and the probability of mistaking $|\Phi_r\rangle$ to $|\Phi_c\rangle$ is $q_{r \to c}$. Thus, the modified detection probability has the same form as Eq. (\ref{e20}), 
\begin{equation}
    (P_c)_{mod}=P_c(1-q_{c \to r})^n+P_rq^n_{ r \to c}.
\end{equation}
According to $|\Phi_c\rangle$ and $|\Phi_r\rangle$ in Eqs. (\ref{e13}) and (\ref{e25}), the corresponding shift of pointer is 
\begin{equation}\label{}
    \begin{aligned}
        &\left \langle \hat{\sigma } _R \right \rangle_c =\frac{\left \langle \Phi_c  |\hat{\sigma } _R |\Phi_c  \right \rangle }{\left \langle \Phi_c  |\Phi_c  \right \rangle} 
        \approx -2ng\frac{\cos n\phi -r\hat{L}}{(1-r\hat{L}\cos n\phi)\sin n\phi}.\\
        &\left \langle \hat{\sigma } _R \right \rangle_r =\frac{\left \langle \Phi_r  |\hat{\sigma } _R |\Phi_r  \right \rangle }{\left \langle \Phi_r  |\Phi_r  \right \rangle} 
        \approx 2ng\frac{p^2\hat{L}\sin n\phi}{(1-r\hat{L}\cos n\phi)(\hat{L}\cos n\phi - r)}.
    \end{aligned} 
\end{equation}
Taking readout errors into account, the real measurement result should be modified to 
\begin{equation}
    \begin{aligned}
        \left \langle \hat{\sigma } _R \right \rangle_m' &= \frac{P_c(1-q_{c \to r})^n\left \langle \hat{\sigma } _R \right \rangle_c+P_rq^n_{ r \to c}\left \langle \hat{\sigma } _R \right \rangle_r}{P_c(1-q_{c \to r})^n+P_rq^n_{ r \to c}} \\
        &=\Gamma'(n)\left \langle \hat{\sigma } _R \right \rangle_c, 
    \end{aligned} 
\end{equation}
where
\begin{equation}
    \Gamma_c(n)= \frac{P_c \left[(1-q_{c \to r})^n-q^n_{ r \to c}\frac{\hat{L}(\hat{L}\cos n\phi - r)}{\cos n\phi - r\hat{L}} \right]}{P_c(1-q_{c \to r})^n+P_rq^n_{ r \to c}} . 
\end{equation}
$\Gamma_c(n)$ is the decrease factor of measurement results of power-recycled scheme. As shown in Figs. \ref{Fig.7}(c)(d), we set $r$, $\gamma$ to appropriate values and draw $\Gamma_c(n)$ changing with $n$ under different postselection distributions and readout errors. Compared to $\Gamma (n)$ of standard scheme, $\Gamma_c (n)$ is generally larger under typical parameters, especially for small entangled qubit numbers. This indicates that recycling can also suppress the influence of readout errors on measurement results, complementing the effect of entanglement. 

Next, we will investigate how power-recycling alters the impact of readout errors on the Fisher information. In practical measurements, we utilize the statistical method of averaging over multiple measurements. Let's assume the $j$th measurement operator is denoted as $\hat{\sigma}_{R,j}$. The probabilities of obtaining the $j$th measurement outcomes from $|\Phi_c\rangle$ and $|\Phi_r\rangle$, respectively, denoted as $w_{c,j}$ and $w_{c,j}$. 
\begin{equation}\label{}
    \begin{aligned}
        &w_{c,j}=\frac{ \langle \Phi_c  |\hat{\sigma } _{R,j} |\Phi_c  \rangle }{\left \langle \Phi_c  |\Phi_c  \right \rangle} 
        \approx -2ng\frac{\cos n\phi -r\hat{L}}{(1-r\hat{L}\cos n\phi)\sin n\phi}.\\
        &w_{r,j} =\frac{ \langle \Phi_r  |\hat{\sigma } _{R,j} |\Phi_r   \rangle }{\left \langle \Phi_r  |\Phi_r  \right \rangle} 
        \approx 2ng\frac{p^2\hat{L}\sin n\phi}{(1-r\hat{L}\cos n\phi)(\hat{L}\cos n\phi - r)}.
    \end{aligned} 
\end{equation}
Thus, the modified probability of observing the $j$th outcome influenced by readout errors is 
\begin{equation}
    h_j=P_c(1-q_{c \to r})^n w_{c,j}+P_rq^n_{ r \to c} w_{r,j}.
\end{equation}
Based on the Fisher information equation $\mathcal{I}(g)=\sum_{j}\frac{(\partial_gh_j)^2 }{h_j}$ \cite{PhysRevA.92.012120}, the modified Fisher information is given by 
\begin{widetext}
    \begin{equation}
        \mathcal{I}_c(g)_{mod}=\frac{\left[ (1-q_{c\to r})^n(w_c^j\partial_gP_c+P_c\partial_gw_c^j)+q^n_{r\to c}(w_r^j\partial_gP_r+P_r\partial_gw_r^j)   \right]^2}{ P_c(1-q_{c \to r})^n w_c^j+P_rq^n_{ r \to c} w_r^j  }.
    \end{equation}
\end{widetext}
When there are no readout errors, $q_{c\to r}=q_{r \to c}=0$, the Fisher information is 
\begin{equation}
    \mathcal{I}_c(g)=\frac{(w_{c,j}\partial_gP_c+P_c\partial_gw_{c,j})^2}{P_cw_{c,j}}.
\end{equation}
So the decrease factor $f_c(n)$ satisfy the following relationship: 
\begin{equation}
    f_c(n)=\frac{\mathcal{I}_c(g)_{mod}}{\mathcal{I}_c(g)}. 
\end{equation}
In figures. \ref{Fig.8}(c)(d), we draw the trend of $f_c(n)$. It can be seen that power-recycling can effectively recover the lost Fisher information, especially in cases with low number entangled qubits. However, as the number of entangled qubits increases, entanglement gradually takes the dominant role, determining the peak and trend of modified Fisher information. Furthermore, the loss rate continues to reduce the Fisher information, especially in scenarios with a large number of entangled qubits. In addition, the introduction of power-recycling has undoubtedly offered a more optimal solution for recovering Fisher information, as evidenced by the increased peak of the curve.

\section{Conclusion}
In summary, we have proposed a power-recycled entanglement-assisted weak measurement model to improve the performance of weak-value-amplification. It solves a key issue in weak measurements: achieving both Heisenberg-scaling measurement and high postselection probability measurement simultaneously. In addition, we found that the recycling component can determine the upper limit of the improvement while the entanglement component can widely enhance the performance among different parameters. 

Furthermore, we considered the effects of walk-off errors and readout errors on the protocol. We found that entanglement broadly increases the walk-off errors, and appropriate parameter selection can offset a significant portion of this effect, leading to improved measurements. 
In addition, we found that the complementary interaction between entanglement and power-recycling further suppressed readout errors, providing a more optimal strategy for recovering the lost Fisher information. 

Power-recycling is not only applicable to enhancing weak measurements assisted by entangled states, but also to  weak measurement of other states such as squeezed states and conherent state \cite{PhysRevLett.115.120401}. Especially, both the pure and mixed coherent states have been demonstrated to achieve Heisenberg-scaling in measuring the distribution of the postselection process \cite{PhysRevLett.114.210801,Jordan2015,PhysRevLett.121.060506,Chen2018}, which matches the recycling technique of modifying this distribution. 

\section{acknowledgements}

This work was supported by the National Natural Science Foundation of China (Grants No. 61875067).
\nocite{*}

\appendix

\end{document}